\documentclass[journal=jacsat,manuscript=article]{achemso}

\usepackage[version=3]{mhchem} 
\usepackage[T1]{fontenc}       
\usepackage{graphics}
\usepackage[usenames, dvipsnames]{color}
\usepackage{longtable}
\usepackage{multirow}
\usepackage{makecell}
\usepackage{natmove}
\setkeys{acs}{usetitle = true}
\usepackage{amsmath,achemso}
\usepackage{bm}
\setkeys{acs}{maxauthors = 9}

\author{Duy Khanh Nguyen}
\affiliation {Department of Physics, National Cheng Kung University, Tainan, Taiwan}

\author{Ngoc Thanh Thuy Tran}
\affiliation {Department of Physics, National Cheng Kung University, Tainan, Taiwan}

\author{Thanh Tien Nguyen}
\affiliation {Department of Physics, College of Natural Sciences, Can Tho University, Can Tho City, Vietnam}

\author{Ming-Fa Lin}
\email{mflin@mail.ncku.edu.tw}
\affiliation {Hi-GEM/Quantum Topology Center, National Cheng Kung University, Tainan, Taiwan}

\title[An \textsf{achemso} demo]
  {Diverse Electronic and Magnetic Properties of Chlorination-Related Graphene Nanoribbons}

\abbreviations{DOS, XPS, STM, ARPES, STS}
\keywords{graphene, halogen, first-principles, chemical bonding, energy gap.}

\begin{document}

\begin{abstract}
The dramatic changes in electronic and magnetic properties are investigated using the first-principles calculations for (Cl, Br, I, At)-adsorbed graphene nanoribbons. The rich and unique features are clearly revealed in the atom-dominated band structures, p-type doping, spin arrangement/magnetic moment, spatial charge distribution, and orbital- and spin-projected density of states. Halogen adsorptions can create the non-magnetic, ferromagnetic or anti-ferromagnetic metals, being mainly determined by concentrations and edge structures. The number of holes per unit cell increases with the adatom concentrations. Furthermore, magnetism becomes nonmagnetic when the adatom concentration is beyond \(60\%\) adsorption. There are many low-lying spin-dependent
van Hove singularities. The diversified properties are attributed to the significant X-C
bonds, the strong X-X bonds, and the adatom- and edge-carbon-induced spin states.

\end{abstract}

\section{I. Introduction}

One-dimensional graphene nanoribbons (1D GNRs) have stirred many experimental\cite{1} and theoretical\cite{2} studies, mainly owing to the honeycomb lattice, nano-scaled thickness, finite-size confinement, edge structures, planar/non-planar structures, and stacking configurations. They are suitable for exploring  the novel physical, chemical and material properties. The 1D GNRs are successfully synthesized by unzipping multi-walled carbon nanotubes\cite{3}, cutting graphene layers\cite{4}, and using other chemical methods\cite{5}. Such systems are expected to have highly potential applications\cite{6} since the essential properties are greatly diversified by the external factors, such as the chemical doping\cite{7,8}, electric field\cite{9}, magnetic field\cite{10,11}, and mechanical strain\cite{12}. This work is mainly focused on the geometric, electronic and magnetic properties of chlorination-related GNRs. Besides, the halogenation effects are investigated thoroughly, in which a detailed comparison between chlorination and fluorination is also made.

Each GNR could be regarded as a finite-size graphene, indicating that its electronic states are sampled from those of the latter  under an open boundary condition\cite{13}.  The low-lying energy bands mainly coming from the $\pi$ bonding of the parallel 2p$_z$ orbitals. According to the theoretical predictions\cite{14} and experimental measurements\cite{15}, it is a direct-gap (\( E_g^{d} \)) semiconductor, and \( E_g^{d} \) strongly depends on the ribbon width. GNRs are predicted to exhibit the spin configurations only near the zigzag edges\cite{16}. Moreover, the semiconductor-metal transitions come to exist in the presence of a transverse electric field \(\vec E_y \) because of the distinct site energies due to the Coulomb potentials\cite{17}.

Electronic and magnetic properties could be dramatically changed by  the chemical modifications. The previous works predict that the alkali adatom adsorptions on GNRs create conduction electrons (n-type doping) and diversify zigzag-edge magnetic configurations\cite{18}. However, fluorinated GNRs present the nonmagnetic semiconducting behavior except that certain low-concentration distributions belong to the p-type doping. This directly reflects the destruction of $\pi$ bondings under very strong F-C bonding. The fluorination effects induce the complicated magnetic configurations\cite{19}. Also, the metallic/semiconducting doping and the magnetic/non-magnetic properties are studied for other adatoms and molecules, e.g., (Co, Ni)\cite{20}, and (CO, NO, NO$_2$, O$_2$, N$_2$, CO$_2$, NH$_3$)\cite{21}. Up to now,  the experimental syntheses and measurements have shown the novel features and the potential applications, covering N-doped GNRs for solar cells under the creation of the efficient metal-free counter electrodes\cite{22}, GNRs and SnO$_2$-nanocluster composite anodes for lithium ion batteries\cite{23}, and hydrogen sensor from Pd-functionalized multi-layer GNR networks\cite{24}.

The  geometric, electronic and magnetic properties of (Cl, Br, I, At)-adsorbed GNRs are investigated using the first-principles method. The binding energies, X-C and C-C bond lengths, adsorption positions, magnetic moments, atom- and spin-dominated  energy bands, spatial charge distributions, free carrier densities, spin configurations, and orbital and spin-decomposed density of states (DOSs) are included in the calculations. The dependence on the  concentrations, adsorption positions and edge structures are explored in detail.  Furthermore, the significant orbital hybridizations in C-C, X-C and X-X bonds (the diverse magnetic configurations) are examined/identified from the atom- and orbital-related properties (the spin-created properties).  This work shows the diversified essential properties under various halogen adatom adsorptions, such as, three kinds of magnetism-dependent metals, the specific relations between halogen concentrations and free hole  densities, the edge-carbon- and halogen-dominated spin distributions, and the halogenation-induced van Hove singularities in DOSs.  The theoretical predictions could be verified by scanning tunneling microscopy (STM)\cite{25}, transmission  electron microscopy (TEM)\cite{26}, angle-resolved photoemission spectroscopy (ARPES)\cite{27},  scanning tunneling spectroscopy (STS)\cite{28,29}, optical spectroscopy\cite{30}, and transport spectroscopy\cite{31}.

\section{II. Computational methods}

The essential properties of chlorination-related  GNRs are explored in detail by the Vienna ab initio simulation package\cite{32} within the spin-polarized density functional theory\cite{33}. The exchange and correlation energies, which come from the many-particle electron-electron interactions, are evaluated from the Perdew-Burke-Ernzerhof functional\cite{34} under the generalized gradient approximation. Furthermore, the projector-augmented wave pseudopotentials can characterize the electron-ion interactions. Plane waves\cite{35}, with a maximum energy cutoff of \(500\)
 eV, are utilized to calculate wave functions and state energies. The 1D periodic boundary condition is along 
\(\hat x\), and the vacuum spacing associated with  \(\hat y\)  and  \(\hat z\)  is larger than \( 15 \)
  \AA  \, to avoid the interactions between two neighboring nanoribbons. The Brillouin zone is sampled by  \( 15 \times 1 \times 1 \)
   and \( 100 \times 1 \times 1 \) k point meshes within the Monkhorst-Pack scheme for geometric optimizations and electronic structures, respectively. The convergence for energy is set to be 
   \(10^{ - 5}   \)
   eV between two simulation steps, and the maximum Hellmann-Feynman force acting on each atom is less than \(  0.01  \)
    eV/\AA \, during the ionic relaxations.  Specifically, the first-principles method could thoroughly evaluate  the various  orbital- and spin-dependent  physical quantities. This is very useful in determining the 
    critical orbital hybridizations of chemical bonds and the diverse spin configurations due to adatoms and/or carbons.

\section{III. Results and discussions}

A systematic study is made on the geometric, electronic and magnetic properties of (Cl, Br, I, At)-adsorbed GNRs under various concentrations and distributions.  Two types of chiral GNRs, with armchair and zigzag edges, are chosen for a model study. AGNR and ZGNR are, respectively, characterized by the number of dimers lines and zigzag lines (N\(_A \)
   and N\(_Z \)) in Figs. 1(a) and 1(b), where the lattice constants along \(\hat x\)  are 
   \(3b\)
     and \( 2\sqrt 3 b \) (\(b\): C-C bond length). The binding energy is calculated by $E_b$ = ($E_{tot}$ - $E_{GNR}$ - n$E_{H}$)/n. The second, third and fourth terms, respectively, correspond to the total energy of the adsorbed system, pristine GNR and isolated halogen adatoms; n represents the number of adatoms. Among four kinds of halogen adatoms, the magnitude of binding energies declines with the increase of atomic numbers, as revealed in Table 1, the Cl-adsorbed GNRs possess the lowest ground state energies. By the detailed examinations, the optimal adsorption position is located at the top site, regardless of any doping concentrations. However, GNRs remain  planar honeycomb structures after the adatom adsorptions, indicating the negligible interactions/bondings between the carbon \( \sigma   \)
      electrons  and  adatom orbitals. The stability of halogen-adsorbed GNRs is decreased with the increase of atomic number. It should be mentioned that the Cl-, Br-, I- and At-absorbed GNRs possess the most stable configurations at higher adatom heights compared to F case, e.g., the chlorinated systems in Table 1.  The optimized X-C bond lengths are shortest in the Cl-adsorbed systems (${\sim\,3.38-3.52}$ \AA \ ). Furthermore, the nearest C-C bond length is lengthened by the  adatom adsorptions ( ${\sim\,1.43-1.6}$ \AA \ ). The variation is about 0.01 \AA \  at low concentrations, while it could reach 0.18 \AA \ under the full adsorption (100 \(\%\)). This indicates an obvious change in the critical chemical bondings after halogenation. 
     
For pristine GNRs, their  band structures are determined by the honeycomb lattice symmetries, quantum confinements and edge structures. The ${N_A=12}$ AGNR exhibits a lot of energy bands initiated from the \(\Gamma \) point, as shown in Fig. 2(a). The occupied valence (v) bands are asymmetric to the unoccupied conduction (c) bands about the Fermi level (\(E_F\)=0), where a direct band gap of \( E_g^d  = 0.6  \)  eV at \(k_x=0\) comes  from the finite-size effect. The  \(\pi\) bonding of the parallel 2p\(_z  \) orbitals  are responsible for the electronic states in the range of \( E^{c,v}  \le 3 \) eV. Furthermore, the \( \sigma   \) bondings due to the (2s, 2p$_x$, 2p$_y$) orbitals dominate the deeper valence states. These are further identified from the orbital-projected DOS (Fig. 6(a)).  Most of the energy bands have the parabolic dispersions, while few of them present the partially flat form. All the energy dispersions vary with wave vector monotonously except for the subband anti-crossings.  The band-edge state, which create the van Hove singularities, appear at \(k_x=\)\(0\) \(\&\) \(1\) (in unit of \(\pi /3b\) ) and other  $k_x$'s related to anti-crossings.

The halogen adsorptions can dramatically change 1D band structures, being very sensitive to adatoms concentrations. As for the single-adatom case [(Cl, Br, I, At) in Figs. 2(b)-2(e)], the valence and conduction states of GNRs present a red shift, in which their energy dispersions almost remain the same after halogenation. The Fermi level is situated in the highest valence band, clearly indicating that the unoccupied states belong to free holes. That is, only this $\pi$-electronic valence band has 1D free carriers. On the other side, the halogen adatoms can induce spin-splitting energy bands with weak/obvious dispersion relations near the Fermi level. Such bands might be partially unoccupied/occupied, leading to the creation of free holes. The total free carrier density is related to the C- and adatom-dominated energy bands crossing $E_F$. Moreover,
 the occupied electron density is non-uniform in the spin-dependent energy bands, so there exists the net magnetic moment, as revealed in Table 1. Cl and Br have the comparable magnetic moment (0.65 ${\mu_B}$ and 0.68 ${\mu_B}$), but larger than that (0.22 ${\mu_B}$ and 0.18 ${\mu_B}$) of I and At, respectively.

The enhanced adatom concentration greatly modifies electronic and magnetic properties. Two carbon-related valence bands become partially unoccupied under two-adatom adsorptions, as shown in Figs. 2(g) and 2(h) for Cl-\((6,\, 21)_s\) and Cl-\((6,\, \textcolor{red}{21})_d\) distributions, in which  Cl-\((6,\, 21)_s\) and Cl-\((6,\, \textcolor{red}{21})_d\) correspond to the adatom distributions under single-side and double-side adsorptions, respectively.  The total free hole density become higher (Table 1), compared with the single-adatom case. Apparently, band structures hardly depend on the single- and double-side distributions.  The almost same results are also revealed in the different positions (Figs. 2(g) and 2(h)). With the further increase of concentration, there are more energy bands with free holes and spin configurations, e.g., the (4Cl)$_d$-adsorption in Fig. 2(i), where (4Cl)$_d$ represent four Cl-doped GNRs under double-side adsorption. But, when the adatom concentration is high enough (${\ge\,60}$ $\%$), the Cl-created spin splitting will disappear, as clearly indicated in Figs. 2(j)-2(l). A lot of Cl-induced conduction and valence bands, with obvious dispersion relations, come to exist in a wide energy range. They are closely related to the great enhancement of the multi-orbital hybridizations in the Cl-Cl bonds (discuss later in charge density distributions).
Such strong bondings can fully suppress the ferromagnetic spin configurations.

 The essential properties are diversified by the boundary structure, since the zigzag-edge carbons create the localized charge distribution and the spin configuration. A pristine ZGNR, as shown in Fig. 3(a),  has a pair of partially flat bands at  \( k_x  < 2/3 \)
   (in unit of  \(  \pi /2\sqrt{3}b\)), corresponding to the edge-localized electronic states. The smallest energy spacing is energy gap, e.g,  \(E_g  = 0.46  \) eV for N\(_Z=8\) ZGNR. They are  doubly degenerate for the spin degree of freedom. The edge-carbon spins will be strongly modified by the halogen adatoms under any concentrations and distributions, and so do  electronic properties and magnetic configurations. As for low concentrations, the energy bands near \(E_F\) are very sensitive to the adatom positions, as indicated in Figs. 3(b)-3(c) and Figs. 3(d)-3(f) for one- and two-adatom adsorptions, respectively. All the low-lying energy bands present the spin splitting, especially for the edge-carbon- and Cl-dominated bands with weak dispersions. This directly reflects the inequivalent magnetic environments of the spin-up and spin-down distributions (Figs. 4(g)-4(l)). When the occupied state densities are not identical in the spin-up and spin-down energy bands, the FM configuration is created by the Cl-adsorption, as observed in Cl-(3), Cl-(14) and Cl-\((3,\, \textcolor{red}{14})_d\) (Figs. 3(b)-3(d)). The opposite is true for two adatoms at distinct zigzag edges (Cl-\((7,\, \textcolor{red}{27})_d\) $\&$ Cl-\((3,\, \textcolor{red}{30})_d\) in Figs. 3(e) and 3(f)).

   Whether there exists a simple relation between the free hole density ( \(\lambda\) ) and the adatom concentration is thoroughly examined for various halogen adsorptions. The former is associated with the partially unoccupied states in carbon- and halogen-dominated valence bands, in which \(\lambda\) is characterized the Fermi momentum under the linear relation \(\lambda  = \frac{2}{\pi }k_F  \). By a lot of numerical calculations (Table  1), the adatom concentration is deduced to play a critical role in determining the following relation: the number of holes per unit cell increases with the adatom concentrations. This is independent of the various distributions, the different kinds of adatoms, and the edge structures. The chlorination-related GNRs might be 1D p-type metals with very high hole  density; therefore, they are expected to be good candidates in electronic devices and electrode materials.

Spin distributions, accompanied with the magnetic moments and the spin-related energy bands, provide much information on magnetic properties. They present the dramatic   transformation during the increase of halogen concentrations, regardless of the edge structures. For example, the ${N_A=12}$ AGNR exhibits the FM configuration under the single-adatom adsorption (Fig. 4(a)), reflecting the Cl-induced magnetic moment (0.65 $\mu_B$ in Table 1). The total range of spin distribution becomes almost double for the two-Cl case (Fig. 4(b)), and so does the net magnetic moment (1.21 $\mu_B$ in Table 1).  However, it grows up with the further increase of concentration, as revealed in the four- and eight-adatom adsorptions (Figs. 4(c) and 4(d)). And then, at the sufficiently high concentration, the strength of the FM configuration declines quickly, e.g., the ten- and thirteen-adatom cases (Figs. 4(e) and 4(f)). Moreover, magnetic configurations disappear when the adatom concentration is beyond the fourteen-adatoms adsorption. Such results clearly illustrate the strong competitions between the FM configuration and the concentration-enhanced X-X bonds. That is, whether the spin configuration could survive is mainly determined by the spin-dependent on-site Coulomb interactions\cite{36} and the adsorption-induced orbital hybridizations.

The spatial spin distributions are greatly diversified by the edge structures. The important differences between ZGNRs and AGNRs lie in the existence of the edge-C spin configuration. Each pristine ZGNR exhibits the FM configuration along two zigzag edges and the AFM one across the ribbon center, as clearly indicated in Fig. 4(g). After chlorination, magnetic configurations strongly depend on the adatom distribution and concentration.  When a single adatom is located at the top site of the upper boundary (e.g., Cl-(3) in Fig. 4(h)), it can induce the spin-up configuration and largely suppresses the spin-up and spin-down distributions due to the neighboring carbons (pristine system in
Fig. 4(g)). The FM configuration is changed into the AFM configuration for two adatoms at different boundaries, such as Cl-\((3,\, \textcolor{red}{30})_d\) in Fig. 4(i). The net magnetic moment is vanishing (Table 1), since the Cl- or the edge-C-induced spin configurations are opposite at two boundaries. The spin distribution keeps similar under two adatoms close to zigzag edges (Cl-\((7,\, \textcolor{red}{27})_d\) in Fig. 4(j)). On the other side, the edge spin configuration hardly depends on the central adatom, e.g., the Cl-(14) adsorption in Fig. 4(k). This adatom has a strong effect on the neighboring spin distribution, so that it creates the almost identical magnetic moment ($\sim$ 0.52 $\mu_B$ in Table 1). Especially, there exist the spin-up dominated FM configurations if two-adatoms are distributed at  edge and center (Cl-\((3,\, \textcolor{red}{14})_d\) in Fig. 4(l)). Cl-\((3,\, \textcolor{red}{14})_d\) adsorption creates the largest magnetic moment (0.65 $\mu_B$ in Table 1) among other cases. The diverse spin distributions directly reflects the magnetic  competitions between the zigzag-edge carbons and adatoms.

The spatial charge distributions, which can account for the rich electronic and magnetic properties, clearly present the critical orbital hybridizations in chemical bonds. They are characterized by charge density ($\rho$) and the variation of charge density after adsorption (${\Delta\rho}$), where the latter is the density difference between X-adsorbed and pristine GNRs. A pristine GNR possesses the $\sigma$ and $\pi$ bondings due to the  planar (2s, 2p$_x$, 2p$_y$) and parallel 2p$_z$orbitals, respectively, as  indicated by the black and red rectangles in Fig. 5(a). The former belong to  very strong covalent bonds with high charge density between two neighboring carbon atoms, so they remain almost the same after the halogen adsorptions (Figs. 5(b), 5(d), 5(f); 5(h)). The latter could survive in a planar GNR for distinct concentrations, distributions and edge structures, being the main mechanism of the metallic behavior. The changes on the $\pi$ bondings are identified through the charge variations between carbon
and chloride atoms (the dashed red rectangles), and they are relatively easily observed at high concentrations (Figs. 5(e) and 5(g)). These  illustrate the single-orbital 3p$_z$-2p$_z$ hybridization in the Cl-C bond. Specifically, there exist  the strong orbital hybridizations in Cl-Cl bonds at the high-concentration adsorptions. The bonding evidences, which are associated with the 3p$_z$-3p$_z$ and (3p$_x$, 3p$_y$)-(3p$_x$, 3p$_y$) orbital hybridizations, are clearly presented in the  charge density difference between two neighboring chlorines (the dashed pink and blue rectangles in  Figs. 5(e) and Fig. 5(g)).

 The orbital- and spin-projected DOSs, as shown in Fig. 6, directly reflect the halogenation-enriched band structures. They further illustrate the orbital hybridizations and magnetic configurations. A lot of asymmetric and symmetric peaks, respectively, arise from the 1D parabolic and weakly dispersive bands. A pristine AGNR has a very wide DOS range due to the 2p$_z$ orbitals (red curves in Fig. 6(a)); that is, the $\pi$ bonding dominates the low-lying peak structure and makes significant contributions to the deeper-energy DOS. The peak structures associated with the $\sigma$ bonding of C-(2p$_x$,2p$_y$) orbitals come to exist at \(E <  - 2.5\) eV (green curves). For the low chlorination, energy spacing between the highest valence peak and the lowest conduction peak present a red shift, indicating the semiconductor-metal transition (Figs. 6(b) and 6(c)). The spin-up and spin-down components are different in the low-energy range of ($-1.5$ eV\(< E <\)$0.5$ eV), being dominated by the 3p$_z$ and (3p$_x$, 3p$_y$) orbitals of Cl adatoms (the dashed pink and blue curves). This indicates the Cl-created FM configuration, as observed in the spin distributions in Figs. 4(a) and 4(b). The spin degeneracy is recovered at high concentrations, e.g., (14Cl)$_d$, (20Cl)$_d$ and (24Cl)$_d$ (Figs. 6(d)-6(f)). Furthermore, the Cl-related DOS covers a very wide range of ($-5$ eV\(< E <\)$3$ eV). These are induced by the strong Cl-Cl bondings. Specifically, the peak structures of the Cl-3p$_z$ and C-2p$_z$ orbitals  appear at the same energies, indicating the 3p$_z$-2p$_z$ orbital hybridization in Cl-C bonds. On the other side, the low-lying DOS peak structures of ZGNRs are very sensitive to the adatom positions even at low concentrations. A pristine ZGNR has a pair of edge-C-dominated symmetric peaks across the Fermi level in the absence of spin splitting (blue circles in Fig. 6(g)). They become the spin-splitting peaks after chlorination, as shown in Figs. 6(h)-6(l). The Cl-dominated peaks exhibit the similar structures. The total occupied states related to these two kinds of peak structures could determine a specific magnetic configuration. The different numbers (same number) of spin-up and spin-down peaks at ${E<0}$ corresponds to the FM (AFM) configuration, e.g., Cl-(3), Cl-(14) and Cl-\((3,\, \textcolor{red}{14})_d\) cases in Figs. 6(h)-6(j) (Cl-\((7,\, \textcolor{red}{27})_d\) and Cl-\((3,\, \textcolor{red}{30})_d\) cases in Figs. 6(k) and 6(l)).

  The chlorination-related and fluorinated GNRs quite differ from each other in the essential properties, being attributed to the very strong orbital hybridizations in the F-C bonds. The fluorination effects cover the shortest F-F bond length (${\sim\,1.5}$ \AA) , the largest binding energy (${\sim\,-3.015}$ eV for a single-F case), the drastic changes of energy bands for any adsorptions (e.g., the F-(14) case in Fig. 2(f)), the p-type metals only under certain low-concentration distributions, and five kinds of electronic and magnetic properties\cite{18}. The F-adsorbed GNRs might be FM and NM metals, NM semiconductors, and AFM semiconductors with/without spin splitting. However, the chlorination-related GNRs belong to the FM, NM and AFM metals, in which the third kind only presents the negligible spin-splitting behavior in zigzag systems (energy bands in Fig. 3(e) and 3(f) and DOS in Fig. 6(k) and 6(l)).

The halogenation-induced diverse phenomena are worthy of a systematic investigation from the experimental measurements, being useful in understanding the critical orbital hybridizations and spin configurations. The STM/TEM measurements on GNRs have confirmed the achiral/chiral open boundaries structures, the nano-scaled widths and the layered structures\cite{37}. They are available in identifying the planar honeycomb lattice after halogen adsorptions, the top-site position, and the X-C bond lengths. As to the high-resolution ARPES, the 1D parabolic energy dispersions, accompanied with an energy gap and distinct energy spacings, are verified for pristine systems\cite{38,39}. Such measurements, with the spin polarization, could be utilized to examine the metallic band structures, the carbon- and adatom-dominated energy bands near the Fermi level, and the spin-split electronic states. The 1D asymmetric peaks of DOSs in semiconducting GNRs are identified from the STS measurements\cite{40}. The  distinct van Hove singularities, being characterized by FM, AFM and NM metals, require further spin-polarized STS examinations. The transport experiments\cite{41} could test  the relation between hole density and halogen concentration, and the spin-polarized currents. 
Moreover, the magneto-optical experiments\cite{42} are suitable for verifying the main features of the spin-split energy bands.

\section{IV. Concluding remarks}

 The detailed first-principles calculations show that the diverse electronic and magnetic properties arise from the critical orbital hybridizations in X-C and X-X bonds and the spin configurations due to Cl adatoms and zigzag-edge C atoms. Such mechanisms are examined/identified from the atom-dominated band structures, free hole densities,  spatial charge densities, net magnetic moments, spin distributions, and  orbital- and spin-projected DOSs.  The predicted chlorination effects on the optimal geometries, energy bands, and van Hove singularities in DOS could be verified using STM/TEM, ARPES and STS measurements, respectively. The further development of the theoretical framework is expected to be suitable for studying emergent layered materials, such as, silicene\cite{43,44}, germanene\cite{45,46} and MoS\(_2\)\cite{47} nanoribbons with the buckled structures, the multi-orbital hybridizations, and the spin-orbital couplings.

  The chlorination-related GNRs present the top-site optimal adsorption positions with the long X-C bond lengths (${3.38-3.52}$ \AA\ ). All of them belong to the p-type metals with FM/NM/AFM configurations, and the number of holes per unit cell increases with the adatom concentrations. The very high concentrations lead to the destruction of the Cl-induced FM configuration and thus  vanishing the magnetic moment. Furthermore, the strong Cl-Cl bonding can create the highly dispersive energy bands, being directly reflected in a wide-range DOS. The distinct edge structures play critical roles in the diversified properties. As for ZGNRs, the carbon and Cl-dominated electronic structures are very sensitive to the adatom positions even under very low concentrations, especially for the low-lying energy bands. Specifically, they  exhibit the FM and AFM spin distributions. The unusual energy dispersions, with/without spin splitting, induce a lot of van Hove singularities in DOS.  The feature-rich essential properties mainly come from the significant orbital hybridizations in X-C, X-X $\&$ C-C bonds, and the spin configurations due to the adatoms and zigzag-edge C atoms. The important differences between chlorination-related and F-adsorbed effects cover the adatom-carbon bond lengths, planar/buckled honeycomb lattices, strength of orbital hybridizations,  metallic or semiconducting behaviors, adsorption-dependent hole densities, changes of energy bands, and  adatom-induced spin states.

  \par\noindent {\bf Acknowledgments}
            
            This research is funded by Vietnam National Foundation for Science and Technology Development (NAFOSTED) under the grant number 103.01-2017.74. Also, we would like to thank  the Physics Division, National Center for Theoretical Sciences (South), the Nation Science Council of Taiwan for the support, under the grant No. NSC 102-2112-M-006-007-MY3.
\bibliography{achemso}

\begin{table}[htb]
                    \caption{Binding energy, magnetic moment/magnetism, energy gap/metal, number of holes in a unit cell, X-C and C-C bond length for N\(_A=12\) armchair and N\(_Z=8\) zigzag GNRs under single- and double-side adsorptions. NM, FM and AFM correspond to non-magnetism, ferro-magnetism and anti-ferro-magnetism, respectively.}
                         \label{table2}
                           \begin{center}
                                                
                           \begin{tabular}{ |l|l|l|l|l|l|l|l|}
                                \hline
                                                GNRs & \makecell{Adsorption\\ configurations }  & \(E_b\) (eV)
                                                 & \makecell{Magnetic\\ moment\\(\(\mu _B \) )/\\magnetism } & \makecell{ \( E_g^{d(i)} \) \\(eV)/ \\Metal}  & \makecell{Number\\of\\holes} & \makecell{X-C\\ (\AA)}  &  \makecell{Nearest\\C-C\\(\AA)}  \\ 
                                                \hline
                                                \multirow{21}{4em}{AGNR\\N\(_A=12\)} & Pristine & \(0\) & 0/NM &  \( E_g^{d}=0.6 \)& 0 & 0 & 1.42 \\ 
                                                 & Cl-\((1)\) & \(-0.6727\) & 0.66/FM & M & 0.95 & 3.38 & 1.43 \\ 
                                                 & Cl-\((14)\) & \(-0.6947\) & 0.65/FM & M & 0.96 & 3.38 & 1.43 \\ 
                                                 & Cl-\((6, 21)_s\) & \(-0.5897\) & 1.21/FM & M & 1.07 & 3.38
                                                 
                                                  & 1.43 \\ 
                                                 & Cl-\((6,\textcolor{red}{21})_d\) & \(-0.6045\) & 1.21/FM & M & 1.07 & 3.38 & 1.43 \\ 
                                          
                                                 & (C:\textcolor{red}{Cl}=24:\textcolor{red}{4})\(_d\) & \(-0.7827\) & 2.43/FM & M & 1.97& 3.38 & 1.43 \\ 
                                               
                                                & (C:\textcolor{red}{Cl}=24:\textcolor{red}{8})\(_d\) & \(-0.8179\) & 5.38/FM & M & 3.86 & 3.40 & 1.48 \\ 
                                                                  
                                              
                                              & (C:\textcolor{red}{Cl}=24:\textcolor{red}{10})\(_d\) & \(-0.8094\) & 2.67/FM & M & 4.85 & 3.40 & 1.48 \\ 
                                                 & (C:\textcolor{red}{Cl}=24:\textcolor{red}{13})\(_d\) & \(-0.8196\) & 0.28/FM & M & 6.46 & 3.50 & 1.56 \\ 
                                                 & (C:\textcolor{red}{Cl}=24:\textcolor{red}{14})\(_d\)& \(-0.8276\) & 0/NM & M & 7.04 & 3.50 & 1.56 \\ 
                                                 
                                                 & (C:\textcolor{red}{Cl}=24:\textcolor{red}{16})\(_d\)& \(-0.8193\) & 0/NM & M & 8.14 & 3.50 & 1.56 \\ 
                                                 & (C:\textcolor{red}{Cl}=24:\textcolor{red}{18})\(_d\)& \(-0.7997\) & 0/NM & M & 9.12 & 3.50 & 1.56 \\ 
                                                 & (C:\textcolor{red}{Cl}=24:\textcolor{red}{20})\(_d\) & \(-0.8143\) & 0/NM & M & 10.05 & 3.50 & 1.56 \\
                                                  & (C:\textcolor{red}{Cl}=24:\textcolor{red}{24})\(_d\) & \(-0.8027\) & 0/NM & M & 12.22 & 3.52 & 1.60 \\ 
                                                  
                                                 & (C:\textcolor{red}{Cl}=24:\textcolor{red}{24})\(_d\)  & \(-0.4074\) & 0/NM & \( E_g^{d}=1.8 \) & 0 & 1.76 &  1.74
                                                 
                                                  \\ 
                                                 & Br-\((14)\)  & \(-0.4893\) & 0.68/FM & M & 0.83 & 3.69 & 1.43 \\ 
                                                 & Br-\((6,\textcolor{red}{21})_d\) & \(-0.3992\) & 1.44/FM & M & 1.05 & 3.69 & 1.43 \\ 
                                                 & (C:\textcolor{red}{Br}=24:\textcolor{red}{14})\(_d\) & \(-0.4059\) & 0/NM & M & 7.06 & 3.70 & 1.50 \\ 
                                                 & (C:\textcolor{red}{Br}=24:\textcolor{red}{24})\(_d\) & \(-0.3962\) & 0/NM & M & 12.06 & 3.70 & 1.50 \\ 
                                                 & I-\((14)\) & \(-0.3189\) & 0.22/FM & M & 0.85 & 4.02 & 1.43 \\ 
                                                 
                                                          & I-\((6,\textcolor{red}{21})_d\) & \(-0.3016\) & 0.44/FM & M & 1.06 & 4.02 & 1.43 \\ 
                                                & (C:\textcolor{red}{I}=24:\textcolor{red}{14})\(_d\) & \(-0.3014\) & 0/NM & M & 7.02 & 4.05 & 1.50 \\ 
                                                
                                                       & (C:\textcolor{red}{I}=24:\textcolor{red}{24})\(_d\) & \(-0.2996\) & 0/NM & M & 12.04 & 4.05 & 1.50 \\

                                                 & At-\((14)\) & \(-0.2750\) & 0.18/FM & M & 0.86 & 4.17 & 1.43 \\ 
                                                \hline
                                                \multirow{6}{4em}{ZGNR\\N\(_Z=8\)} & Pristine & \(0\) & 0/AFM & \( E_g^{d}=0.46 \) & 0 & 0 & 1.42 \\ 
                                                 & Cl-\((3)\) & \(-1.0095\) & 0.26/FM & M & 0.95 & 3.12 & 1.43 \\ 
                                                 & Cl-\((14)\) & \(-1.0216\) & 0.52/FM & M & 0.94 & 3.12 & 1.43 \\ 
                                                 & Cl-\((3,\textcolor{red}{14})_d\) & \(-0.7943\) & 0.65/FM & M & 1.28 & 3.12 & 1.43 \\ 
                                                  
                                                 & Cl-\((7,\textcolor{red}{27})_d\) & \(-0.8138\) & 0/AFM & M & 1.82
                                                  & 3.12 & 1.43 \\ 
                                                 & Cl-\((3,\textcolor{red}{30})_d\) & \(-0.9656\) & 0/AFM & M & 1.98 & 3.12 & 1.43 \\ 
                                                 & Br-\((14)\) & \(-0.7605\) & 0.58/FM & M & 0.92 & 3.62 & 1.43 \\ 
                                              
                                                 & I-\((14)\) & \(-0.5038\) & 0.65/FM & M & 0.82 & 3.90 & 1.43 \\ 
                                             
                                                 & At-\((14)\) & \(-0.3688\) & 0.70/FM & M & 0.76 & 4.08 & 1.43 \\ 
                                                \hline
                                                \end{tabular}
                                                 \end{center}
                                                 \end{table}

\newpage
\begin{figure}[!h]
\centering
\includegraphics[width=9cm, height=20cm]{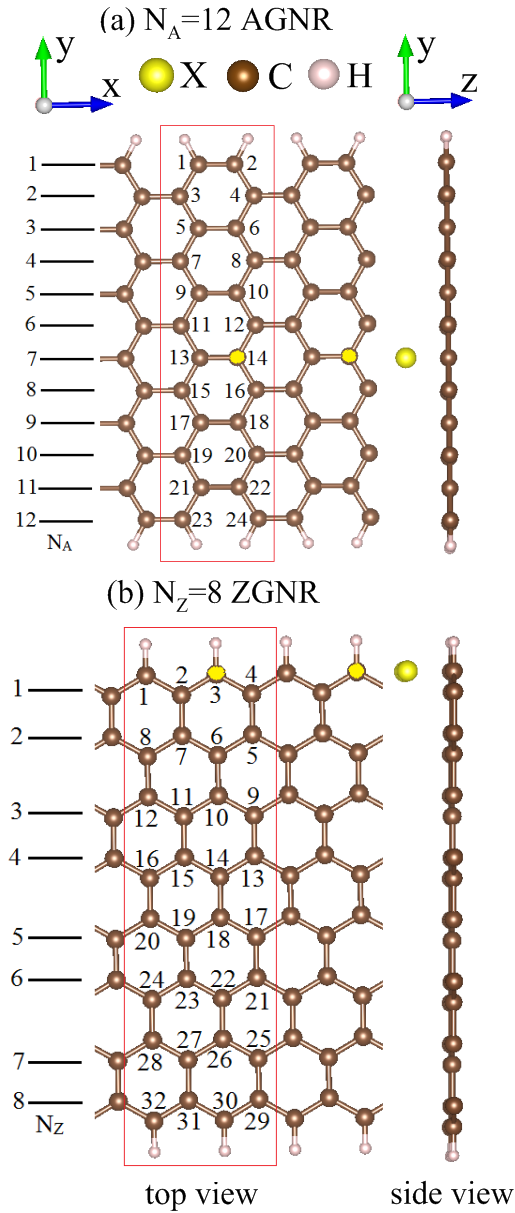}
\caption{The top- and side-view geometric structures of halogen-adsorbed GNRs for (a) \(N_A\)=12 armchair and (b) \(N_Z\)=8 zigzag systems. The red rectangles represent unit cells used in the calculations. The top-site positions of adatom adsorptions are characterized by the numbers on carbon atoms.}
\end{figure}

\newpage
\begin{figure}[!h]
\centering
\includegraphics[width=7.5cm, height=20cm]{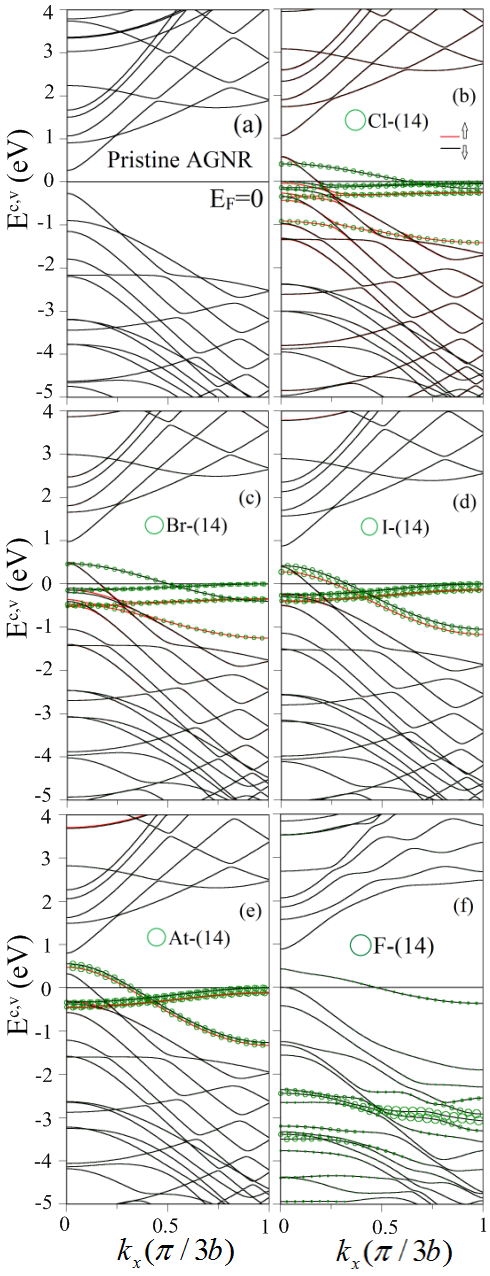}
\end{figure}

\newpage
\begin{figure}[!h]
\centering
\includegraphics[width=7.5cm, height=20cm]{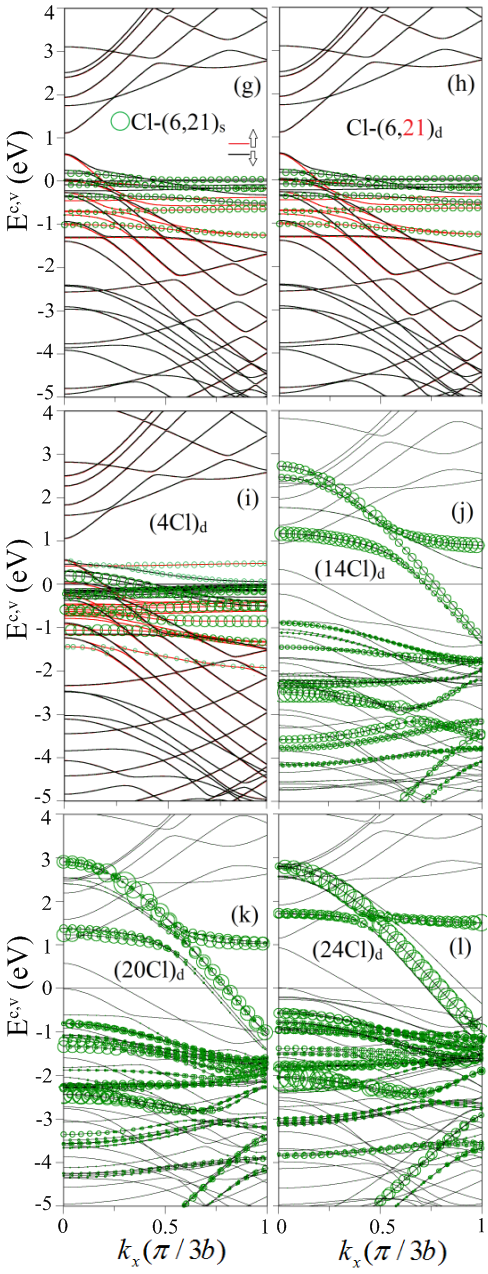}
\caption{Band structures of (a) pristine, (b) Cl-(14)-, (c) Br-(14)-, (d) I-(14)-, (e) At-(14)-, (f) F- (14)-,  (g) Cl-\((6, 21)_s\)-, (h) Cl-\((6,\textcolor{red}{21})_d\)-, (i) (4Cl)\(_d\)-,  (j) (14Cl)\(_d\)-, (k) (20Cl)\(_d\)- and (l) (24Cl)\(_d\)-adsorbed \(N_A\)=12 AGNRs. Green circles represent the contribution of halogen adatoms. The red and black curves, respectively, correspond to the spin-up and spin-down energy bands.}
\end{figure}

\newpage
\begin{figure}[!h]
\centering
\includegraphics[width=7.5cm, height=20cm]{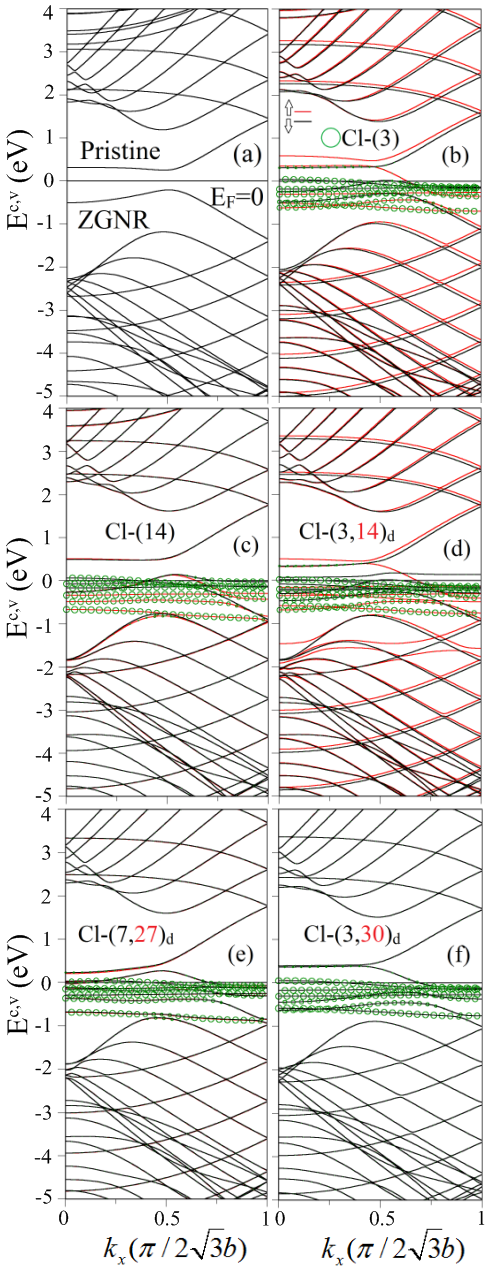}
\caption{Similar plot as Fig. 2, but shown for (a) pristine, (b) Cl-(3)-, (c) Cl-(14)-, (d) Cl-\((3,\textcolor{red}{14})_d\)-, (e) Cl-\((7,\textcolor{red}{27})_d\)- and (f) Cl-\((3,\textcolor{red}{30})_d\)-adsorbed \(N_Z\)=8 ZGNRs. }
\end{figure}

\newpage
\begin{figure}[!h]
\centering
\includegraphics[width=15cm, height=16cm]{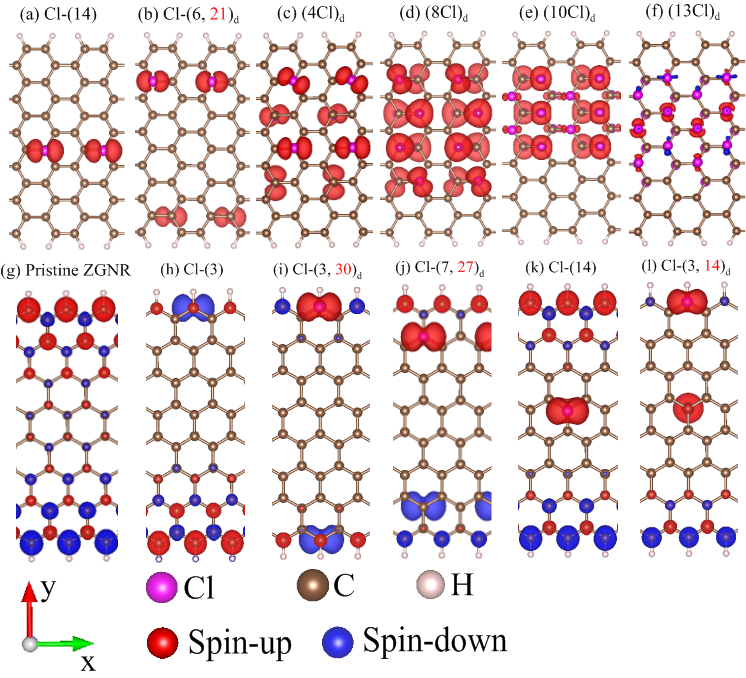}
\caption{The spatial spin densities for (a) Cl-(14)-, (b)Cl-\((6,\textcolor{red}{21})_d\)-,  (c) (4Cl)\(_d\)-, (d) (8Cl)\(_d\)-, (e) (10Cl)\(_d\)-, (f) (13Cl)\(_d\)-adsorbed \(N_A\)=12 AGNRs,  and for (g) pristine, (h) Cl-(3)-,  (i) Cl-\((3,\textcolor{red}{30})_d\)-, (j) Cl-\((7,\textcolor{red}{27})_d\)-, (k) Cl-(14)-, (l) Cl-\((3,\textcolor{red}{14})_d\)-adsorbed \(N_Z\)=8 ZGNRs.}
\end{figure} 
\newpage
\begin{figure}[htb]
                \includegraphics[width=10cm, height=20cm]{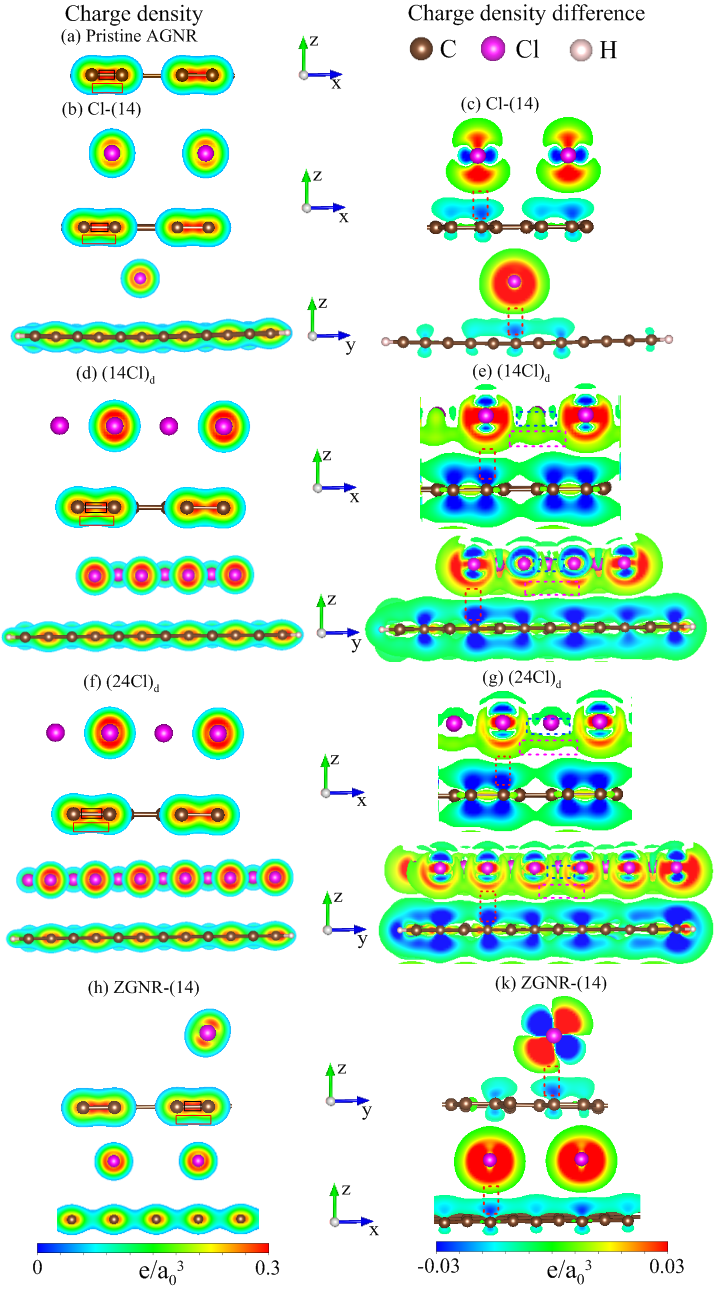}
                \caption{Spatial charge density situate at the left-hand side for (a) pristine, (b) Cl-(14)-, (d) (14Cl)\(_d\)-, (f) (24Cl)\(_d\)-adsorbed \(N_A\)=12 AGNRs, and  (h) ZGNR-Cl-(14); charge density difference situate at the right-hand side  for (c) Cl-(14)-, (e) (14Cl)\(_d\)-, (g) (24Cl)\(_d\)-adsorbed \(N_A\)=12 AGNRs, and  (k) ZGNR-Cl-(14). $a_0$ is Bohr radius.}
                \label{fgr:6}
              \end{figure}

\newpage
\begin{figure}[htb]
                          \includegraphics[width=9.5cm, height=20.5cm]{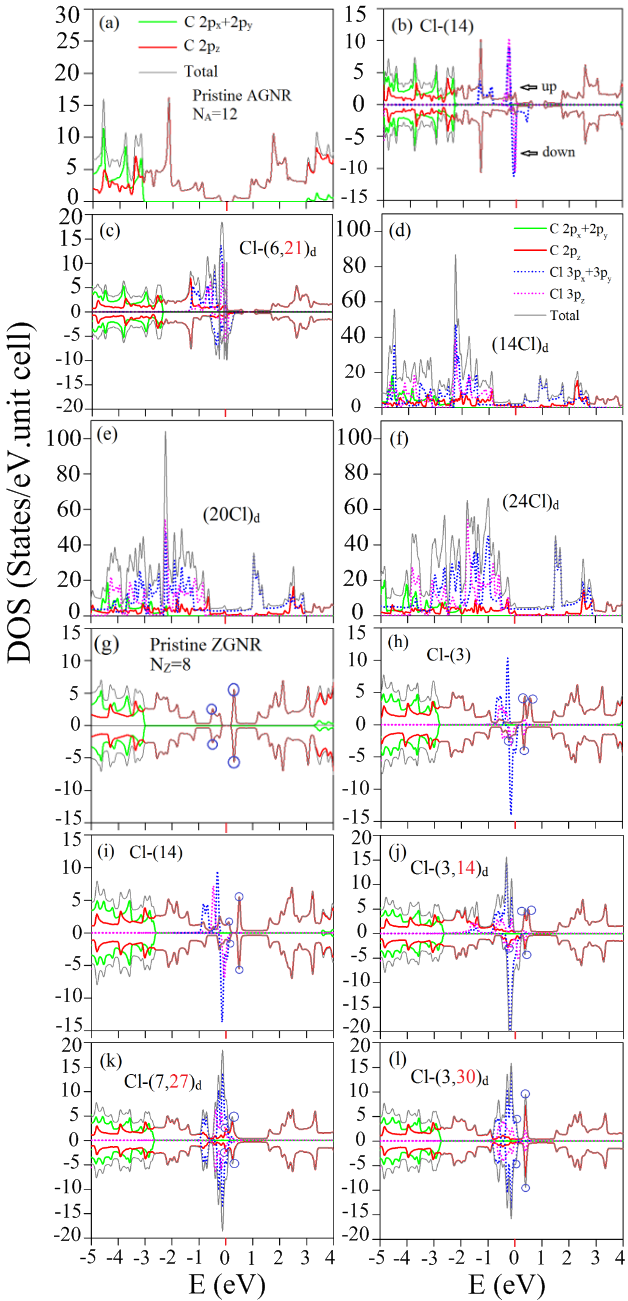}
                                     \caption{Orbital- and spin-projected  DOSs for (a) pristine, (b) Cl-(14)-, (c) Cl-\((6,\textcolor{red}{21})_d\)-, (d) (14Cl)\(_d\)-, (e) (20Cl)\(_d\)-, (f) (24Cl)\(_d\)-adsorbed \(N_A\)=12 AGNRs;  (g) pristine,  (h) Cl-(3)-, (i) Cl-(14)-, (j) Cl-\((3,\textcolor{red}{14})_d\)-, (k) Cl-\((7,\textcolor{red}{27})_d\)-, (l) Cl-\((3,\textcolor{red}{30})_d\)-adsorbed \(N_Z\)=8 ZGNRs. Blue circles correspond to the partially flat bands. The spin-up and spin-down components are also shown for the magnetic systems.}
                                     \label{fgr:7}
                               \end{figure} 
\end{document}